\documentclass[preprint2]{aastex}

\begin{document}

\title{A Deep {\it ROSAT} HRI Observation of  NGC 1313}

\author{Eric M. Schlegel\altaffilmark{1}, Robert
Petre\altaffilmark{2}, E. J. M. Colbert\altaffilmark{2,3}, Scott Miller\altaffilmark{3}}

\altaffiltext{1}{High Energy Astrophysics Division, Smithsonian Astrophysical Observatory, Cambridge, MA  02138}

\altaffiltext{2}{X-ray Astrophysics Group, Laboratory for High Energy Astrophysics, NASA-GSFC, Greenbelt, MD 20771}

\altaffiltext{3}{Department of Astronomy, University of Maryland, College Park, MD 20742}

\begin{abstract}

We describe a series of observations of NGC 1313 using the {\it ROSAT}
HRI with a combined exposure time of 183.5 ksec.  The observations
span an interval between 1992 and 1998; the purpose of observations
since 1994 was to monitor the X-ray flux of SN1978K, one of several
luminous sources in the galaxy.  No diffuse emission is detected in
the galaxy to a level of $\sim$1-2$\times$10$^{37}$ ergs s$^{-1}$
arcmin$^{-2}$.  A total of eight sources are detected in the summed
image within the D$_{25}$ diameter of the galaxy.  The luminosities of
five of the eight range from $\sim$6$\times$10$^{37}$ to
$\sim$6$\times$10$^{38}$ erg s$^{-1}$; these sources are most likely
accreting X-ray binaries, similar to sources obseved in M31 and M33.
The remaining three sources all emit above 10$^{39}$ erg s$^{-1}$.  We
present light curves of the five brightest sources.  Variability is
detected at the 99.9\% level from four of these.  We identify one of
the sources as an NGC 1313 counterpart of a Galactic X-ray source.
The light curve, though crudely sampled, most closely resembles that
of a Galactic black hole candidate such as GX339-4, but with
considerably higher peak X-ray luminosity.  An additional seven
sources lie outside of the D$_{25}$ diameter and are either foreground
stars or background AGN.

\end{abstract}


\section{Introduction}

One of the most significant discoveries using the {\it Einstein}
Observatory was the existence in many nearby spiral galaxies of X-ray
sources with luminosities in the range $\sim$10$^{39-40}$ erg
s$^{-1}$.  As the brightest Galactic X-ray binary has an X-ray
luminosity of a few $\times$10$^{38}$ erg s$^{-1}$, these {\it
Einstein} sources represented a new class.  Some of these sources are
located in the central regions of spirals (e.g., NGC 6946, IC 342, and
the northern source in NGC 1313 \citep{FT87}; M33 \citep{MR83}), so
they were originally believed to be low-activity active nuclei or
starburst regions.  Others, however, are well-separated from the
galactic nucleus (e.g., the southern source in NGC 1313; \citealt{FT87})
and could not be so easily explained.

With the more powerful X-ray observatory {\it ROSAT},
higher-sensitivity observations of many nearby galaxies have been
possible.  In NGC 6946 \citep{Sch94} and IC 342 \citep{Breg93}, the
luminous nuclear regions have been resolved into an ensemble of
discrete sources.  In NGC 1313, the supposed nuclear source was shown
to be located a full kiloparsec from the optical nucleus
\citep{Co+95, CM99}.  While spectroscopic observations using
the X-ray observatory {\it ASCA} have shown that the central source in
M33 is not an active nucleus \citep{Tak94}, the nature of the NGC
1313 ``central'' source remains undetermined.

NGC 1313 is a barred spiral galaxy (SBd; \citet{deV+91}) in the
southern hemisphere.  Its distance is estimated to be 4.5 Mpc
\citep{deV63}.  Numerous bright star-forming regions are present in
the galaxy \citep{Ryd+95}.  The only very luminous X-ray
source optically identified is SN1978K, the first supernova
identified as such on the basis of its X-ray emission \citep{Ryd93}.

Using a set of {\it ROSAT} HRI observations obtained over the past 6
years, we have detected two additional highly luminous sources within
1 kpc of the nucleus.  We find that all three central sources exhibit
temporal variability, providing the first evidence that they are all
compact.  We also present the summed image of all of the individual
observations which provides a deep look at NGC 1313.

\section{Observations}

Nine {\it ROSAT} HRI observations have been obtained over the years
1992-1998.  Eight were collected as part of a
monitoring program of SN1978K \citep{SPC96, Sch+99}; the
ninth was obtained to pursue the optical identification of a different
source \citep{Sto94, Sto95}.  The summary of the complete data
set is contained in Table~\ref{comp_set}.  A single {\it ROSAT}
observation was typically spread across several calendar weeks;
the MJD listed in the table is the value near the center of the
observing window.

\begin{figure*} 
\caption{Stacked image of NGC 1313 from nine separate epochs of {\it
ROSAT} HRI observations.  The total exposure time is $\sim$182 ksec at
the center of the image.  Source labels generally lie to the right of
the source, except in the crowded region around the nucleus.  Sources
are listed in Table~\ref{src_list} in order of decreasing RA.  Source
11 is SN1978K.  Several of the weaker sources are circled for clarity.
The largest circle schematically defines the D$_{25}$ diameter
($\sim$9$'$).}
\includegraphics{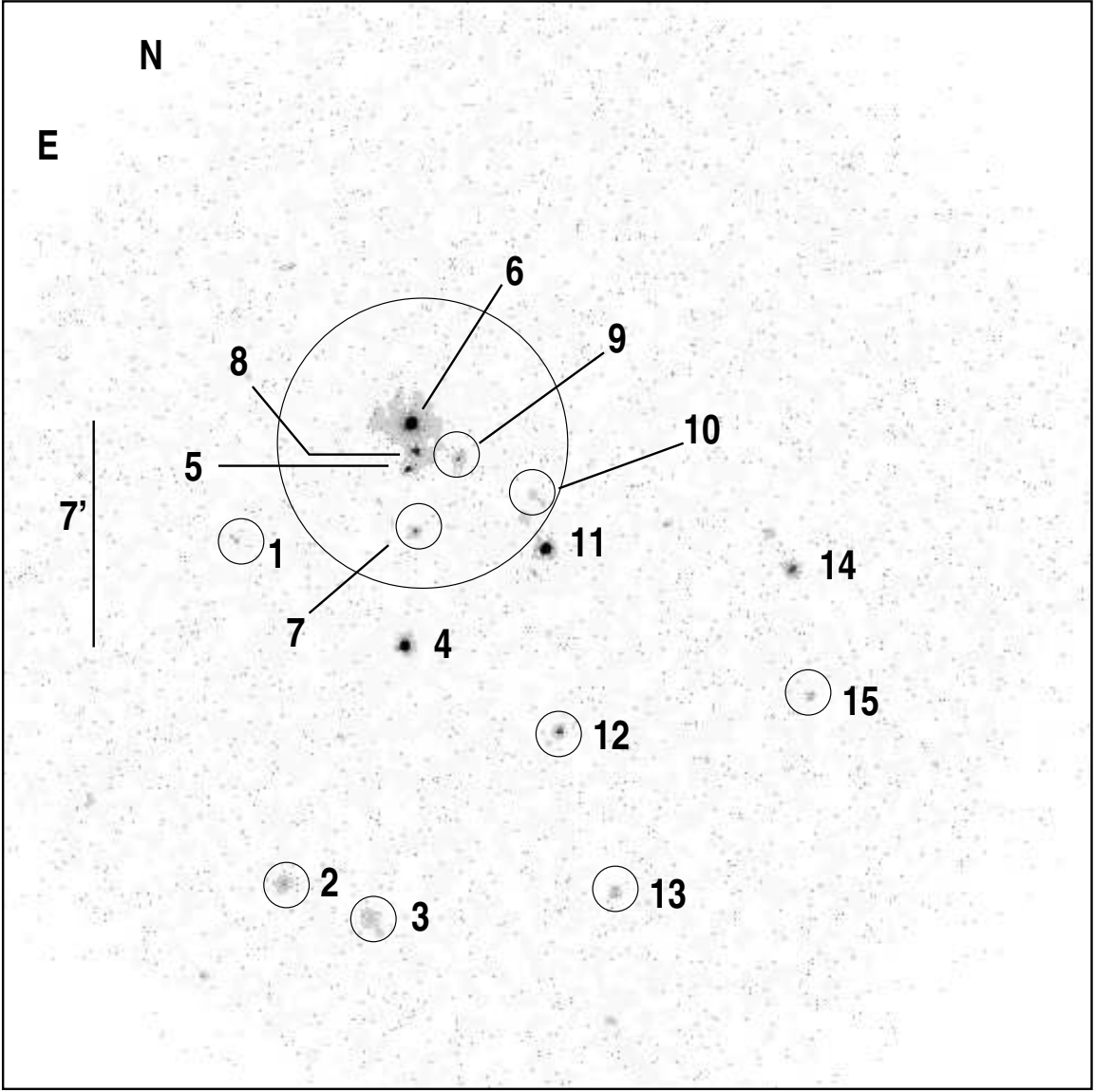}
\label{stacked}
\end{figure*}

The data have all been reduced identically.  Each observation was
processed with the ``extended source'' software described by
\citet{Sno98} and available at the US {\it ROSAT} Science Data Center
at NASA-GSFC.  The software accounts for the particle background,
filtering out times of anomalously high background and other bad
quality data.  Each observation was also corrected for the aspect
error described by \cite{Har99}.  This error arose from incorrect
times (the fractional portion) that are associated with each aspect
record which led to incorrect positions for each event.  The data were
affected because the point spread function was broadened by
$\sim$3$''$.  The correction was applied to the first seven
observations; the last two were unaffected by the error (the eighth
and ninth observations had zero or nearly zero fractional times, so
the error was at the level of $\sim$0$''$.1, well below
detectability).  Once the data were filtered and aspect-corrected, the
images from all of the observations were registered (using the
IRAF/STSDAS task ``registration'') and stacked.  Figure~\ref{stacked}
shows the resulting stacked image.

The {\it Chandra} Observatory source detection task ``wavdetect,''
based on the wavelet code of \cite{Mic96}, was used to locate point
sources using a detection criterion of $>$4.5 ${\sigma}$ (at most,
$<$1 false source).  The detection threshold varies with off-axis
angle.  At the center, the threshold is $\sim$2$\times$10$^{-4}$
counts s$^{-1}$ which corresponds to a luminosity of
$\sim$5$\times$10$^{37}$ ergs s$^{-1}$ assuming a thermal spectrum of
temperature 3 keV and absorbed by the known column density toward NGC
1313 (N$_{\rm H}$ $\sim$6$\times$10$^{20}$ cm$^{-2}$ from an E$_{\rm
B-V}$ of 0.11 \citep{SFD98} and using the N$_{\rm H}$-E$_{\rm B-V}$
relation of \citealt{PS95}).  At a radius of 10$'$, the threshold is
$\sim$2.5$\times$10$^{-4}$ counts s$^{-1}$ or $\sim$7$\times$10$^{37}$
ergs s$^{-1}$.

A total of 15 sources were found in the summed image; the detection
task ignored any ``source'' within 2$'$ of the edge of the field where
the vignetting is severe.  The detected sources are listed in
Table~\ref{src_list} with their positions (J2000), total exposure time
at the source's position, detected counts, estimated flux and
luminosity, and PSPC counterpart name \citep{Co+95,Mi98}.  The sources
are ordered by decreasing right ascension.  For this paper, we number
the sources and use the numbers for the source name instead of the
longer but complete {\it ROSAT} source number.

We extracted the counts for each source using an aperture of 40$''$,
taking care to exclude counts from nearby sources.  The background was
extracted from an annulus surrounding the entire galaxy.  The counts
from any source falling within the background annulus were excluded.
The vignetting within $\sim$5$'$ of the center is small ($\sim$5\%)
but the correction was applied to any source with an off-axis angle
$>$1$'$.  No energy filtering was applied during the extraction, so
the extracted counts represent the incident flux for the complete
bandpass of the HRI (0.1--2.4 keV).

The HRI efficiency was relatively constant throughout the {\it ROSAT}
mission.  The HRI high voltage was increased to compensate for a loss
of gain on 1994 June 21.  As a consequence, the HRI flux sensitivity
increased by $\sim$10\%; any sources showing fluctuations at or below
the 10\% level were interpreted as constant.  As we describe below,
the variability of several sources is too large to be interpreted as
resulting from the gain change.  Note, in particular, that all
observations but the first occurred after the gain change.

The luminosity of each source is estimated assuming it is located
within NGC 1313; several of the sources are likely to be foreground
(or background) objects, so the true luminosity will be lower (or
higher).  The unabsorbed luminosities (Table~\ref{src_list}) in the
0.1--2.4 keV band were estimated by comparing the source count rates
with that expected from a source with a thermal spectrum (kT = 2 keV)
and a column density identical to the Galactic column.  For a column
density of 1.5$\times$10$^{21}$ cm$^{-2}$, the approximate mean value
for N$_{\rm H}$ obtained from fitting the PSPC data of sources 6
(=X-1) and 4 (=X-2) \citep{Co+95, Mi98}, the unabsorbed luminosities
decrease by $\sim$25\%.  The luminosities differ by $\sim$25\% if the
adopted temperature is instead set to 7 keV.  The luminosities also
differ by $\sim$10-15\% assuming other adopted thermal models at a
temperature of 2 keV but differ by $\sim$80\% for an adopted power law
model.  The power law is particularly sensitive to the column density
value.

\begin{figure*}
\caption{X-ray contours superposed on the digitized UK Schmidt Sky
Survey image of NGC 1313.  (a) the entire galaxy scaled for high
grayscale values (left) and low grayscale values (right); (b) an
expanded view near the nucleus, scaled to show the inner galaxy.}
\label{opt_xray}
\scalebox{0.5}{\rotatebox{-90}{\includegraphics{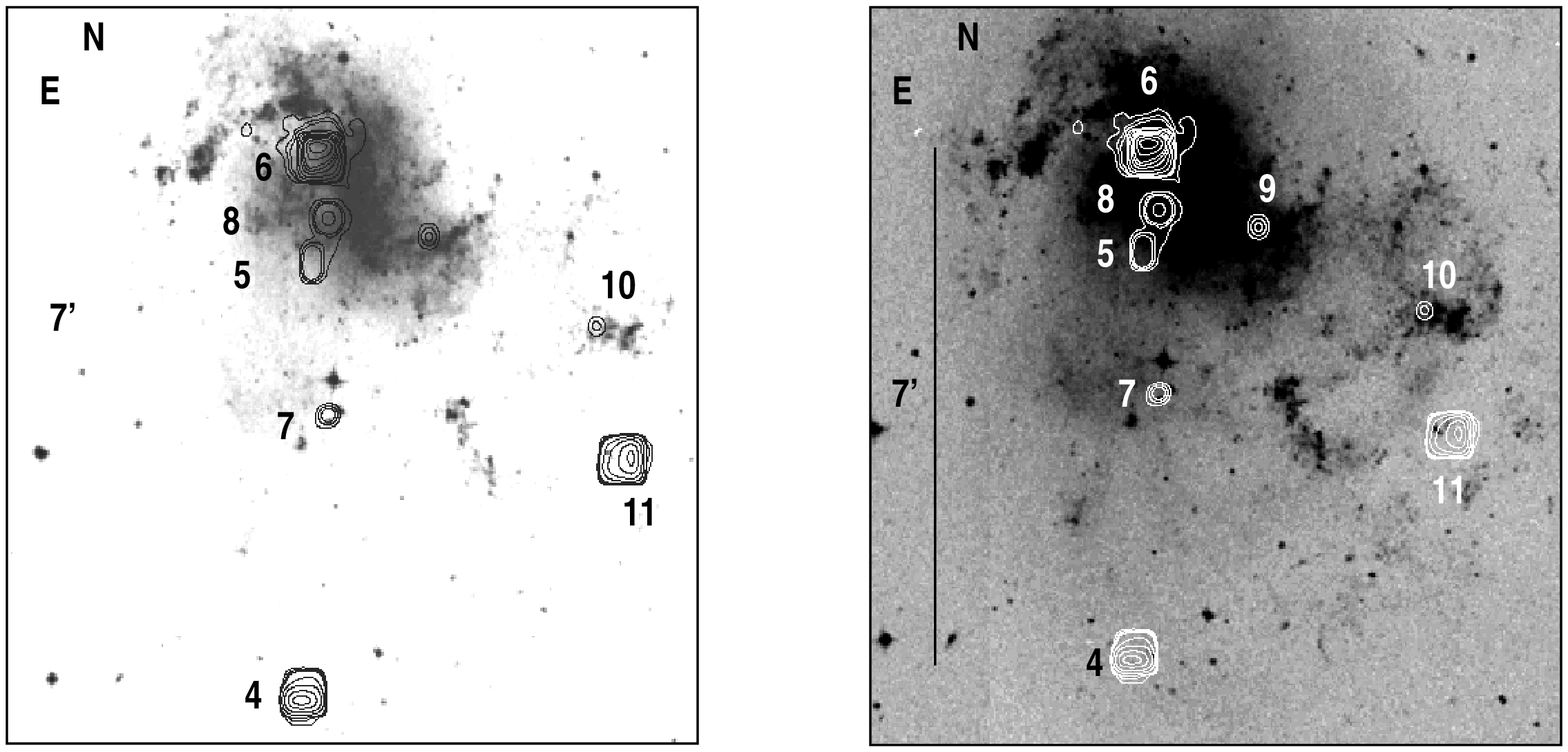}}}
\scalebox{0.5}{\includegraphics{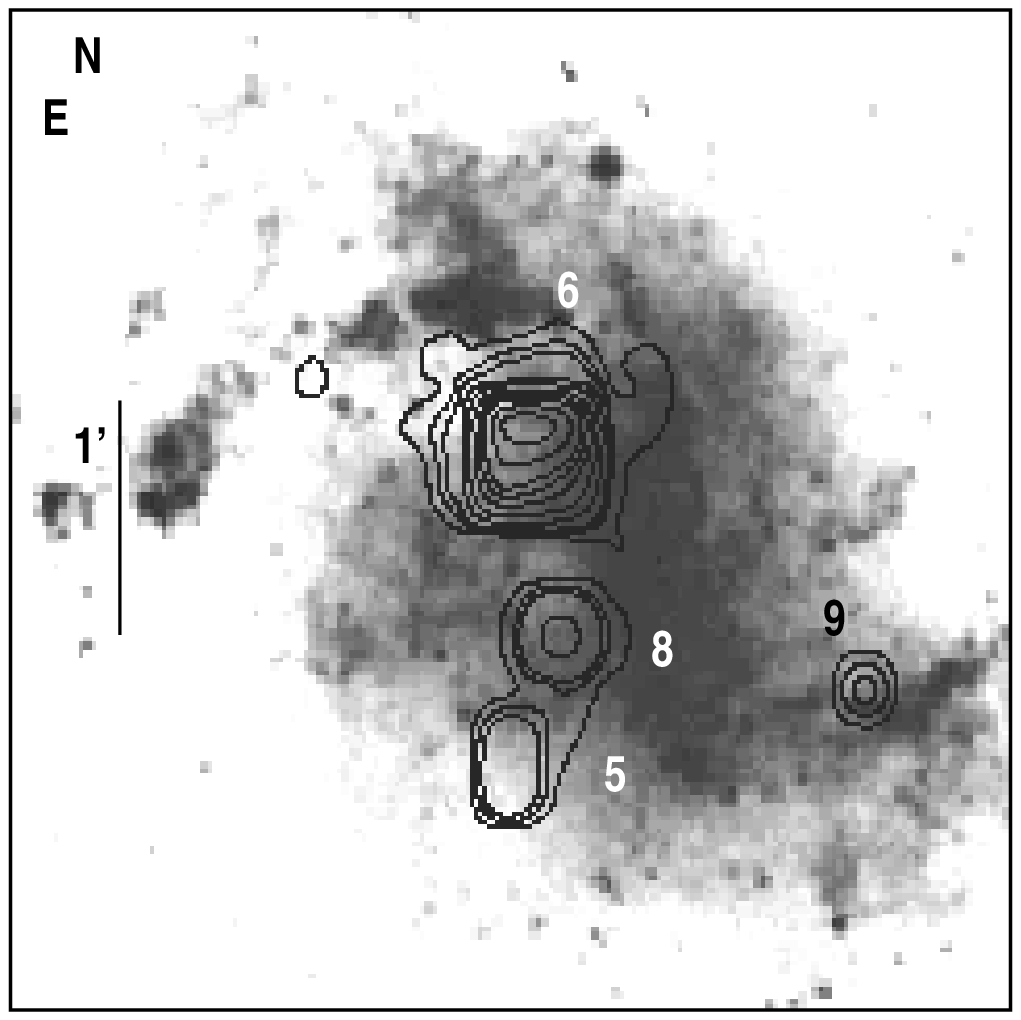}}
\end{figure*}

Figure~\ref{stacked} shows the composite image of NGC 1313.  
Except in the crowded region near the nucleus, the appropriate
Table~\ref{src_list} label is placed directly beneath each source.

\section{The Summed Image}

Figures~\ref{opt_xray}a and \ref{opt_xray}b show X-ray contours
superposed on an optical image\footnote{Image obtained via Skyview
facility at NASA's HEASARC; original digitized from the UK Schmidt
Telescope Southern Sky Survey} of NGC 1313.  The first figure shows
the entire galaxy with the gray scale set at two different values: a
``high'' scaling and a ``low'' scaling.  The second image shows the
center of the galaxy with the contrast set to show the inner region.
None of the sources lie on the bar.  Source 6 falls just east of the
northern end of the bar.  Source 7 lies just east and approximately
midway along the bar.  Source 9 is positioned on the northern side of
the west trailing arm.  Source 5 lies $\sim$1$'$ east of the southern
end of the bar.  While all of the sources lie east of the bar, the
positional match of source 11=SN1978K with its optical position
ensures that the aspect solution is correct.
   
The deep image shows no evidence of diffuse emission across the face
of the galaxy (diffuse emission may be present surrounding source
source 6, which we will discuss shortly).  An adaptively-smoothed
image, generated using a box size requiring a minimum of 99 counts,
showed a circularly-symmetric image that resembled the exposure map.
In other words, the adaptively-smoothed image is consistent with the
expected vignetting gradient.  To assign an upper limit to the
presence of diffuse emission, we extracted the counts in a circular
aperture, centered on the galaxy, with a radius corresponding to half
of its D$_{25}$ diameter ($\sim$9$'$, \citet{Tul88}).  We excluded the
counts for all point sources that fell within the aperture (sources 4
through 11).  As a check on our procedure, we also extracted the blank
region NW and S of NGC 1313.  We corrected these off-axis regions for
the difference in extracted areas and for the reduced exposure time
resulting from vignetting.  We corrected for the vignetting using the
exposure map that is a product of the extended source software of
\citet{Sno98}.

A net total of 297$\pm$243 counts was extracted within the on-galaxy
aperture; the area encompassed on the sky is $\sim$55.8 arcmin$^2$.
For a mean exposure time of 182.6 ksec across this region, the net
counts correspond to a count rate of 16.3$\pm$13.3$\times$10$^{-4}$
cts s$^{-1}$.  Adopting the 3${\sigma}$ upper limit of
$\sim$4$\times$10$^{-3}$ counts s$^{-1}$, the 3${\sigma}$ upper limit
on the luminosity in the 0.1--2.4 keV band is $\sim$1$\times$10$^{39}$
ergs s$^{-1}$ or $\sim$1.5$\times$10$^{37}$ ergs s$^{-1}$
arcmin$^{-2}$ across the extraction region.  This limit is about a
factor of 2 higher than the PSPC limit \citep{Mi98} of
$\sim$7$\times$10$^{36}$ erg s$^{-1}$ cm$^{-2}$ (correcting an
arithmetic error in that paper).  The factor of 2 lies within 30\% of
the expected difference based on a calculation of the instrumental
backgrounds and sensitivities between the PSPC and the HRI.

\begin{figure*} 
\caption{Azimuthally-averaged radial profiles centered on source 6
(top) and 4 (bottom) compared with the {\it ROSAT} HRI calibrated
point spread function from HZ 43 \cite{Dav97}.  The background level
is indicated by the dashed line.  The PSF profile of HZ 43 is included
in the top figure (filled triangles) solely for comparison.  Note the
excess near 15-20$''$ in Source 6; the same behavior is visible in HZ
43 \citep{Dav97}.}
\label{profile} 
\scalebox{0.6}{\includegraphics{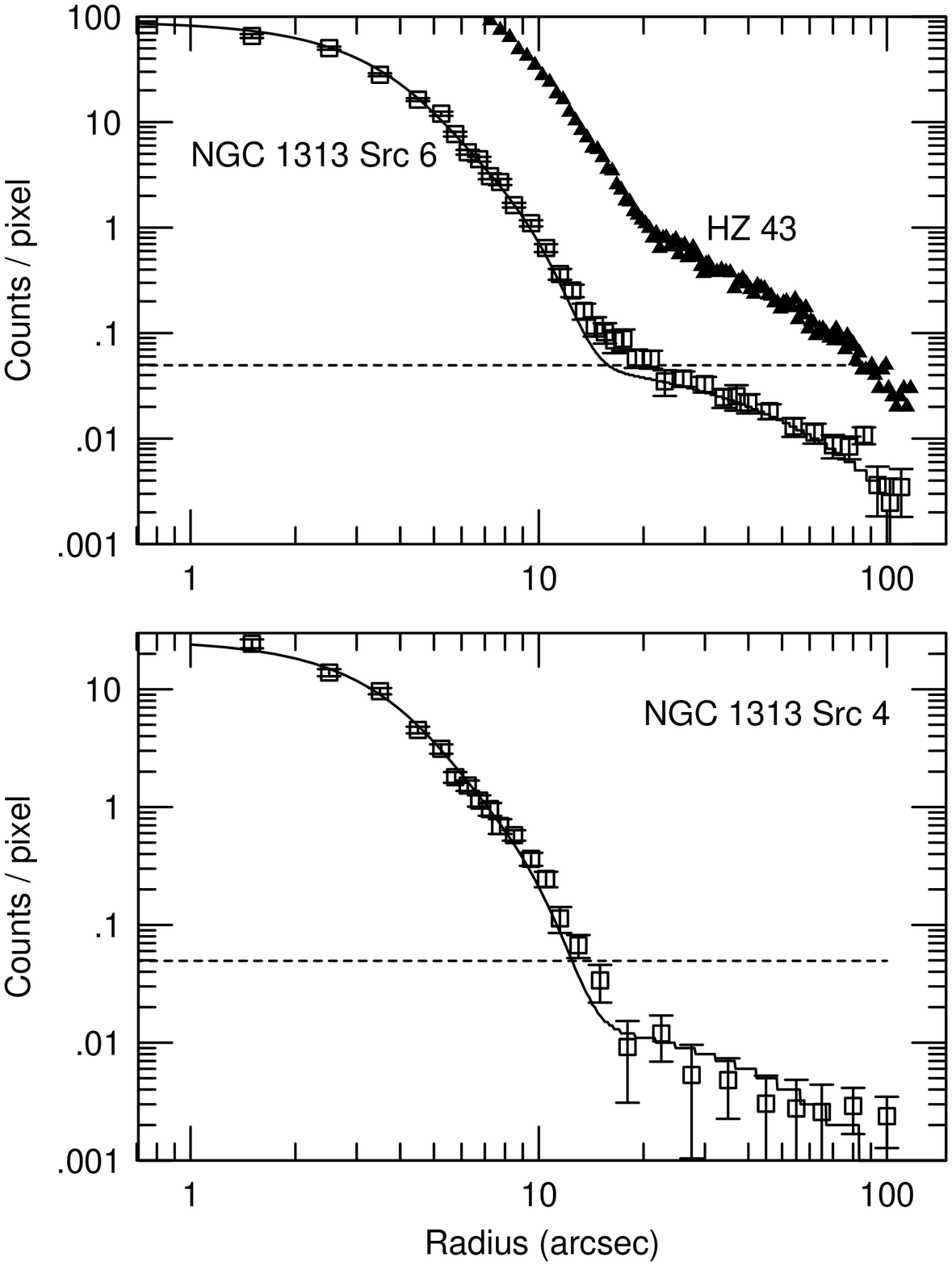}}
\end{figure*}

In Figure~\ref{stacked}, diffuse emission appears to be concentrated
around source 6.  To examine whether this emission is truly diffuse,
we extracted the counts from sources 6 and 4 in annuli centered on
each source and extending outward to $\sim$2$'$.  Extraneous sources
falling within the annuli were excluded.  Source 4 is believed to be a
point source without any extended or nearby emission \citep{Sto94,
Sto95}.  The resulting radial profiles are shown in
Figure~\ref{profile}.  The background counts were extracted from an
annulus surrounding the entire galaxy and are shown as a dashed line
in each figure.  The solid line in each figure shows the calibrated
point spread function \citep{Dav97} using the data from a long
exposure of HZ 43.  We fit the PSF curve to the data in the
5$''$-10$''$ and 30$''$-50$''$ range.  The fit yielded a
${\chi}^2/{\nu}$ of 1.28 for 39 degrees of freedom (the PSF is defined
in \citep{Dav97}, so the fit is for the scale factor) which
corresponds to a significance of $\lesssim$85\% for any deviations from
the model PSF.  A fit to the HZ 43 profile yeilded a similar
significance.

Neither profile shows any evidence for extended emission when compared
to HZ 43.  The slight excess observed at the bend in the PSF at
$\sim$15--18$''$ is also visible in the HZ 43 data \citep{Dav97}; the
same few points near $\sim$18$''$ that lie above the functional form
of the PSF in sources 4 and 6 also do so in the HZ 43 data.  Beyond
about 20$''$, the known HRI halo begins to dominate.  The visual
impression of diffuse emission surrounding source 6 results solely
from the long exposure that reveals the instrumental halo.

The nucleus of NGC 1313 is also not detected.  Its position at
03:18:17.5, -66:29:48 (J2000, \citet{Ryd+95}), corresponds to an
offset of about 0$'$.6 NW of source 8 in Figure~\ref{opt_xray}b.  We
extracted the counts in an aperture centered on the nucleus and
corrected them for the known halo surrounding source 6.  These counts
represent an upper limit on the flux at the nucleus and correspond to
a 3${\sigma}$ count rate of 3.7$\times$10$^{-4}$ counts s$^{-1}$ or a
flux of $\sim$4$\times$10$^{-14}$ ergs s$^{-1}$ cm$^{-2}$.  At the
distance of NGC 1313, the luminosity is $\sim$9$\times$10$^{37}$ ergs
s$^{-1}$ in the 0.1--2.4 keV band.

\section{The Sources}

Before launching into a description of the individual sources, we
first comment on the sources {\it en masse}.  Sources 1, 2, 3, 12, 13,
14, and 15 lie outside of the D$_{25}$ region of NGC 1313 and are very
likely not associated with the galaxy.  We describe them briefly in
the appendix.  From the {\it ROSAT} log N-log S relation \cite{Has94},
we estimate a total of $\sim$4-6 extragalactic sources lie within the
HRI field-of-view.  Within the D$_{25}$ radius, we expect
$\sim$0.2-0.4 extragalactic sources.  Both estimates provide support
for the comment that the seven sources are external to NGC 1313.

For the sources detected by both the PSPC and the HRI, the HRI
positions agree to within $\sim$10$''$ with the PSPC positions
presented in \cite{Co+95, Mi98}.  These positional uncertainties are
typical of PSPC positions.  

The luminosity estimated for source 11 (= SN1978K) agrees with the
luminosity estimate obtained from the PSPC data to better than
$\sim$15\%.  Source 11 is the only approximately constant source in
NGC 1313 during the years of observation.  The luminosity estimates
for sources 4, 6, 10, and 14 differ by factors ranging from
$\sim$0.2-0.3 (sources 4 and 6) to $\sim$0.5 (sources 10 and 14).  We
believe these differences are dominated by the variability of the
sources (described below).  For two of the weaker sources (7 and 8),
the estimates differ by a factor of two.  We estimate the fluxes can
change by $\sim$25\% if a model with an incorrect temperature is
adopted.

The luminosity of source 9 differs from the luminosity measured in the
two PSPC observations by a factor of $\sim$4.  If we estimate the
expected HRI count rate by reducing the PSPC count rate by the PSPC-to-
HRI count conversion factor of $\sim$2.5, we estimate the summed HRI
image should yield 109$\pm$21 counts.  Instead, we detect only
38$\pm$11 counts; the difference is significant at the $>$99\% level
and establishes source 9 as a variable object.  The PSPC data sets
show essentially no differences in the count rates between the two
observations in 1991 and 1993.  The ``variability'' must then be a
long-term change that operates over a scale of years.

We have searched for variability of the sources using the HRI data
sets.  In Table~\ref{src_rates}, we summarize the data and the results
of this search.  For each source we list the count rate or upper limit
for each observation.  For the weaker sources, we have added together
adjacent observations to improve the significance of some of the
detections.  We have performed a chi-squared test against the
hypothesis of constant source flux, using the average detected flux.
By ignoring upper limits, we obtain a chi-squared that represents a
conservative estimate of the likelihood of source variability.  As can
be seen from Table~\ref{src_rates}, sources 4, 5, 6, 8, and 11 all are
variable with high probability ($>$97 percent).

\begin{figure*} 
\caption{HRI light curves covering the nine epochs of observations (a)
sources 4 and 6 (solid lines), and source 11 = SN1978K (as a dashed
line) (top) and source 11 (bottom).  Note that the scale for the
bottom plot covers 1/4 of the range of the top plot.  (b) light curves
for the sources 5 and 8.  The upside-down triangles represent upper
limits. The diamond on the light curve of source 5 is the 90\% upper
limit on the flux of source 5 in the {\it ROSAT} PSPC observation
\cite{Mi98}.}
\label{light_curve} 
\scalebox{0.4}{\includegraphics{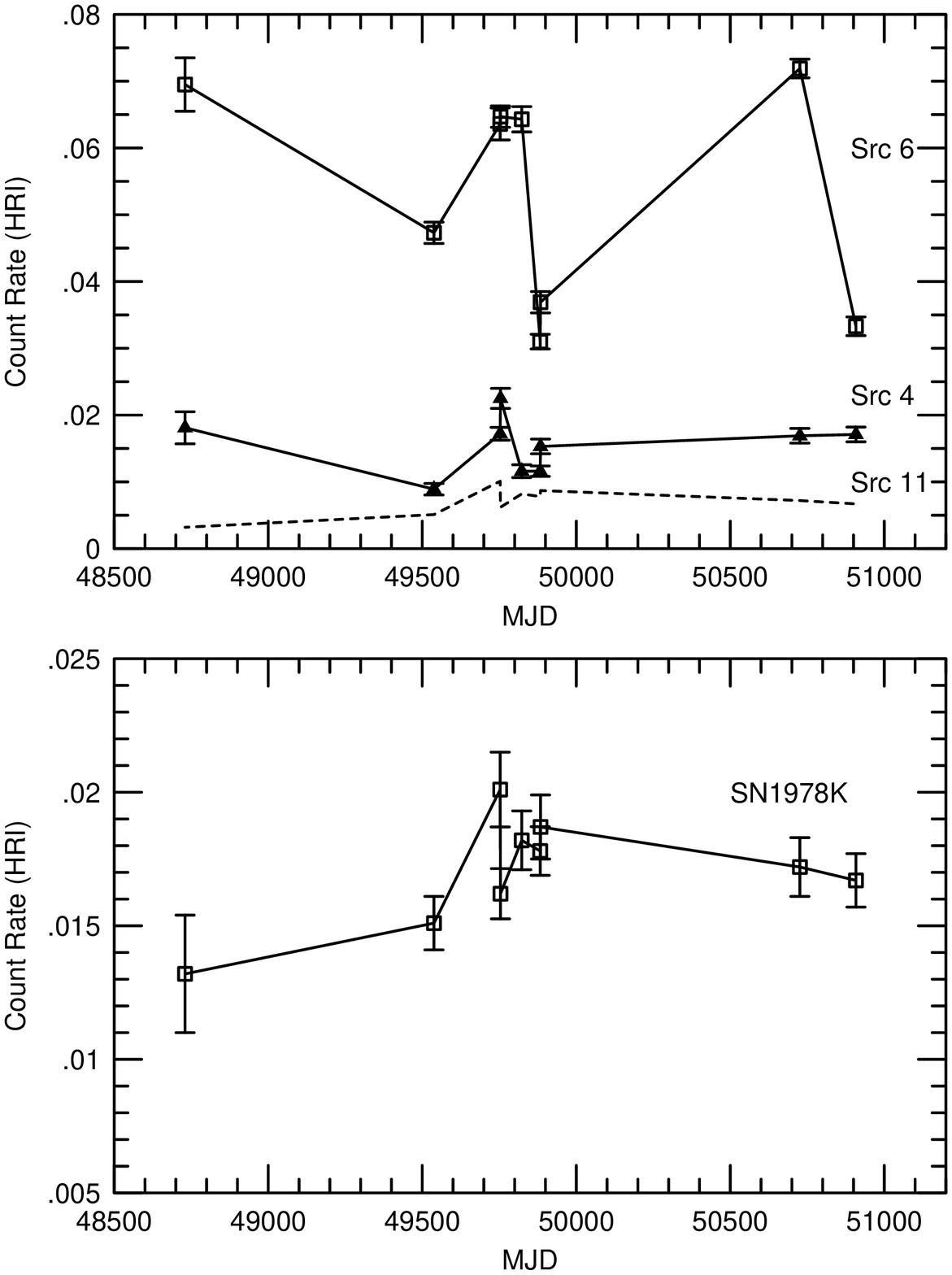}}
\scalebox{0.4}{\includegraphics{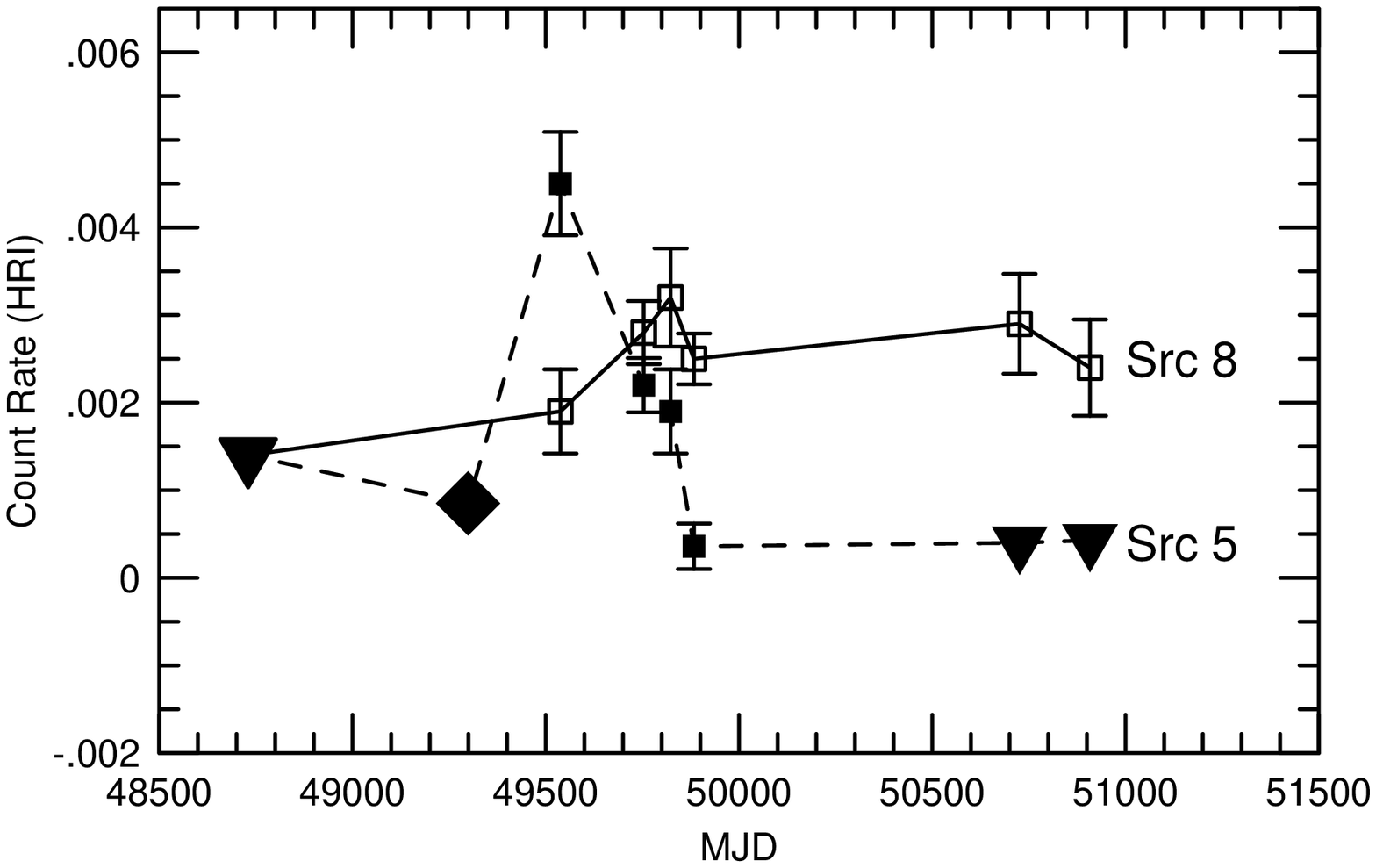}}
\end{figure*}

Source 6 (= X-1 in \citet{Co+95, Mi98}) is located near the northern
end of the bar \citep{Co+95}.  Studies by \citet{Ryd+95} and
\citet{Pet93} indicate that it lies north of the dynamical and
photometric center of NGC 1313 by $\sim$1 kpc ($\sim$40$''$).  There
is no known counterpart, in either the optical or the radio.
\citet{CM99}, in a survey of nearby galaxies, show that about half of
the nearby galaxies have a luminous source located just off of the
dynamical center of the galaxy.  Source 6 is the most luminous source
in the {\it ROSAT} band in NGC 1313, with L$_{\rm X}$ $\sim$2
$\times$10$^{40}$ ergs s$^{-1}$.  It was one of the two sources
detected in the 1980 {\it Einstein} IPC observation \citep{FT87}.
There has been little evidence that it varies other than the two PSPC
observations which were separated by more than two years
\citep{Mi98}. The HRI variability (Figure~\ref{light_curve}, top) is a
crucial piece of evidence on the source size, namely, it must be
compact having undergone a factor of 2 change in $\lesssim$60 days
(near MJD 49850).  From the light travel time, 60 days corresponds to
a physical size $<$0.05 pc.  The variability, luminosity, and spectrum
\citep{Co+95,Ptr94a,Ptr94b} suggest it is an active nucleus-like
source with a mass of $\sim$10$^{3-4}$ M$_{\odot}$ while the lack of a
counterpart and the displacement from the nucleus provide evidence
against this hypothesis and simultaneously testifies to its unusual
nature.

Source 4 (= X-2 in \citet{Co+95}; \citet{Mi98}), the other source
detected in the IPC observation \citep{FT87}, has been thought by some
to be a nearby Galactic source, perhaps a cooling neutron star
\citep{Sto94, Sto95}.  The {\it ASCA} SIS data, however, indicate that
the X-ray spectrum is not that expected for a cooling neutron star (T
$\sim$10$^6$ K), but is more typical of a low-mass X-ray binary: a
thermal spectrum with kT $\sim$ 2 keV \citep{Ptr94a, Ptr94b}.  Unlike
source 6, intensity variability was detected from it between the 1980
IPC and the 1991 PSPC observation \citep{Co+95}.  Colbert et al. argue
that the source is a black hole candidate with a mass of $>$100
M$_{\odot}$ based upon its PSPC + IPC spectrum and its super-Eddington
luminosity.

The X-ray luminosity of SN1978K (= source 11 in this paper = X-3 in
\citet{Co+95, Mi98}); \cite{Ryd93}) is formally variable at the 97\%
confidence level across the 9 HRI observations
(Figure~\ref{light_curve}, bottom).  The largest flux change from the
mean of the 9 observations is $\sim$23\%.  \cite{SPC96}, however,
stated that SN1978K was constant within the errors.  We reconcile
these statements by recalling that the ${\chi}^2$ estimate tests for
variability within the observation set.  For SN1978K, however, we are
also interested in whether a trend exists for the entire data set.
Essentially, the 97\% variability can be considered as the
``micro-variability;'' the long-term trend would then be the
``macro-variability.''  The details of the fit to the light curve are
described in \cite{Sch+99}.  Very briefly, the X-ray luminosity should
decay as L $\propto$ t$^{\alpha}$.  A fit to the data gives ${\alpha}$
= (-3.71$\pm$6.52)$\times$10$^{-5}$ (90\% confidence range).  In other
words, there is no decay, so the long-term luminosity is constant
within the errors.  The luminosity did increase by at least a factor
of 5 from 1980, when it was not detected by the {\it Einstein} IPC
(1.5 years after optical maximum) to its discovery in 1991 with the
{\it ROSAT} PSPC \citep{Ryd+95, SPC96}.  We refer readers to the cited
papers for details.

One other supernova has appeared in NGC 1313: SN1962M.  At the
position of SN1962M (03:18:15,-66:29:51 J2000), we extracted
29.5$\pm$11.6 counts which corresponds to a 3${\sigma}$ flux of
$\sim$2.5$\times$10$^{-14}$ ergs s$^{-1}$ cm$^{-2}$ or a corresponding
luminosity upper limit of $\sim$5$\times$10$^{37}$ ergs s$^{-1}$.
There is no evidence that SN1962M has been detected.

The remaining sources may be accreting X-ray binaries similar to
sources detected in M31 and M33.  Source 8 appears to be a relatively
steady X-ray source, perhaps a persistent X-ray binary with a
luminosity of $\sim$6$\times$10$^{38}$ ergs s$^{-1}$
(Figure~\ref{light_curve}b).  \cite{Mi98} showed that the hardness
ratio, defined as (H-S)/(H+S), with H = photons with energies $>$1 keV
and S = photons with energies $<$1 keV, is +0.26.  For a source with
that hardness absorbed by the known Galactic column density, a thermal
bremsstrahlung spectrum must have kT $\gtrsim$2 keV.

Source 5 may be the most dramatic source (Figure~\ref{light_curve}b).
No source was detected in the first HRI observation, but the exposure
time was sufficiently short that the upper limit is not particularly
restrictive.  The second HRI exposure shows an easily detected source.
This exposure occurred after the gain change.  The increase in flux of
source 5 is not only substantially larger than the $\sim$10\% change
in sensitivity, but it also is not matched by the other sources.
Sources 4 and 6 became considerably fainter in the second observation;
sources 7 and 11 brightened slightly (and perhaps consistently with
the increased sensitivity from the gain change).  Subsequent exposures
show source 5 fading by about a factor of 10 in $\sim$300 days.  When
the full light curve is accumulated, it is easy to conclude that an
outburst occurred near the time of the second exposure.  The increase
in flux is at least a factor of 3, from L$_{\rm X}$
$\sim$3$\times$10$^{38}$ to $\sim$10$^{39}$ ergs s$^{-1}$.  We also
included the PSPC observation described in \citet{Mi98} because it
falls near the time of brightening.  No PSPC source is detected at the
location of source 5; the plotted flux is the 90\% flux upper limit
for source 5.  The flux was converted to an HRI count rate assuming a
thermal bremsstrahlung spectrum absorbed by the Galactic column to NGC
1313.

Most X-ray transients do not reach L$_{\rm X}$ $\sim$10$^{39}$ erg
s$^{-1}$ levels unless the source is an X-ray binary (e.g.,
\citealt{WNP95}).  A supernova could produce the light curve.  The
presence of SN1978K guarantees considerable scrutiny of NGC 1313.
Observations were made at MJD 49361 (day -177) and 49722 (day +184) in
the optical; radio observations were obtained on MJD 49284 (day -254)
and 49627 (day +89) \citep{Sch+99}.  We judge it unlikely a supernova
could explode and fail to be reported.  In the discussion to follow,
we assume source 5 is a binary X-ray transient.

To compare the source's outburst to the X-ray light curves of known
Galactic X-ray sources, we examined the 2-10 keV light curves obtained
with the {\it RXTE} All-Sky Monitor.  Ideally, a comparison should be
made using X-ray binaries observed with the HRI.  Few have been
observed because of the high column density toward the Galactic
center, and none monitored over a duration comparable to the NGC 1313
observations.  The known column density to NGC 1313 is relatively low,
as measured in the PSPC spectrum of source 6 \citep{Co+95, Mi98}.  On
the basis of the column density, we estimate the mean energy of the
source 5 photons to be $\gtrsim$1.2 keV, so the ASM light curves
provide an approximate comparison to the HRI band.  A precise
comparison is not the goal; we are interested in whether any of the
known Galactic X-ray transients have light curve shapes similar to
that of source 5.

\begin{figure*}
\caption{NGC 1313 source 5 HRI light curve compared to Galactic X-ray
source light curves for XTE J1550-56, GX339-4, GRS1915+105, X0115+634,
Aql X-1, and X1705-44.  The data points all belong to source 5.  Upper
limits are shown by triangles with apex pointing down.  The diamond is
the PSPC observation from \cite{Mi98}.  The light curves of the X-ray
binaries have been smoothed (with a 5-point triangular response)
solely for clarity.}
\label{xray_nova}
\scalebox{0.4}{\includegraphics{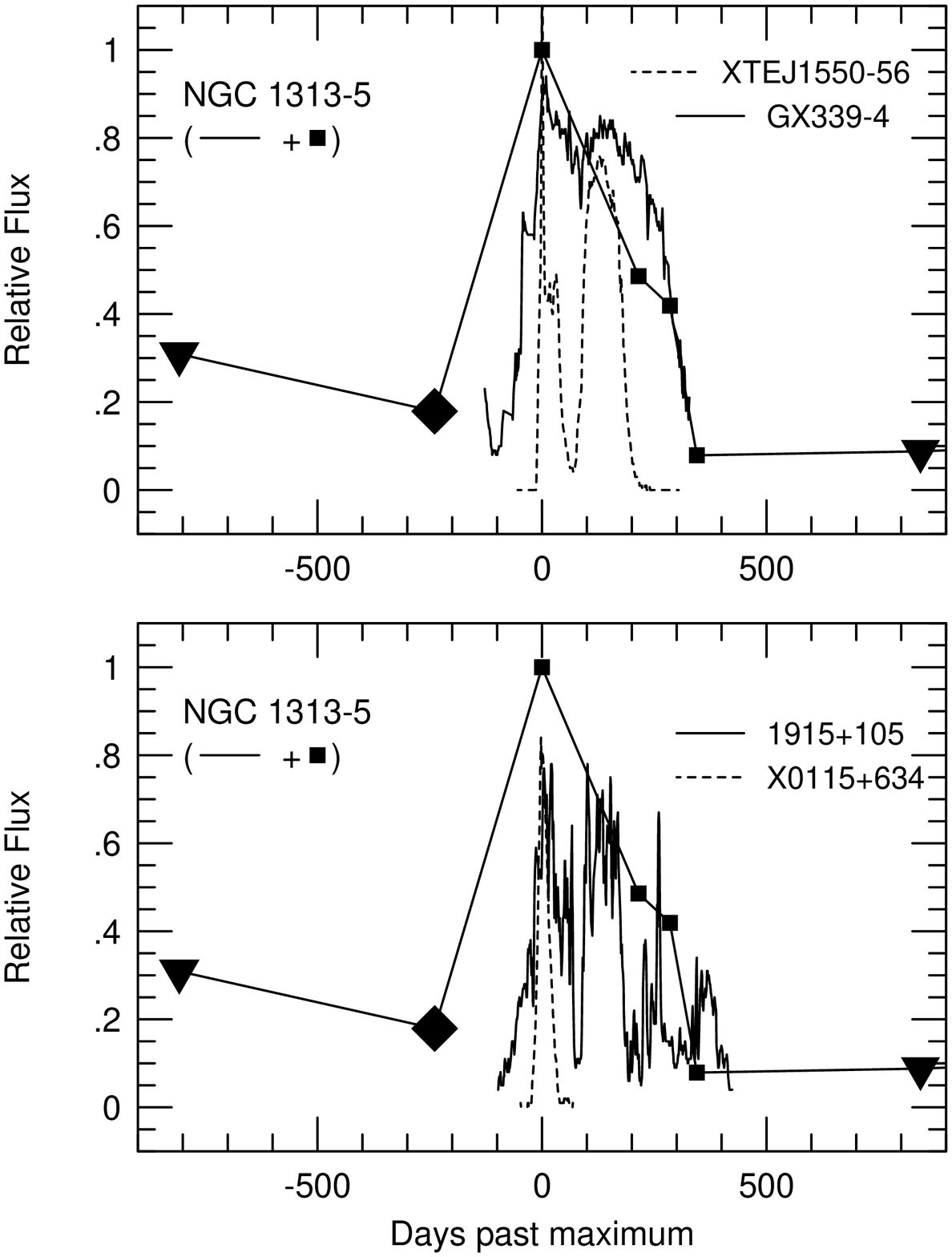}}
\scalebox{0.4}{\includegraphics{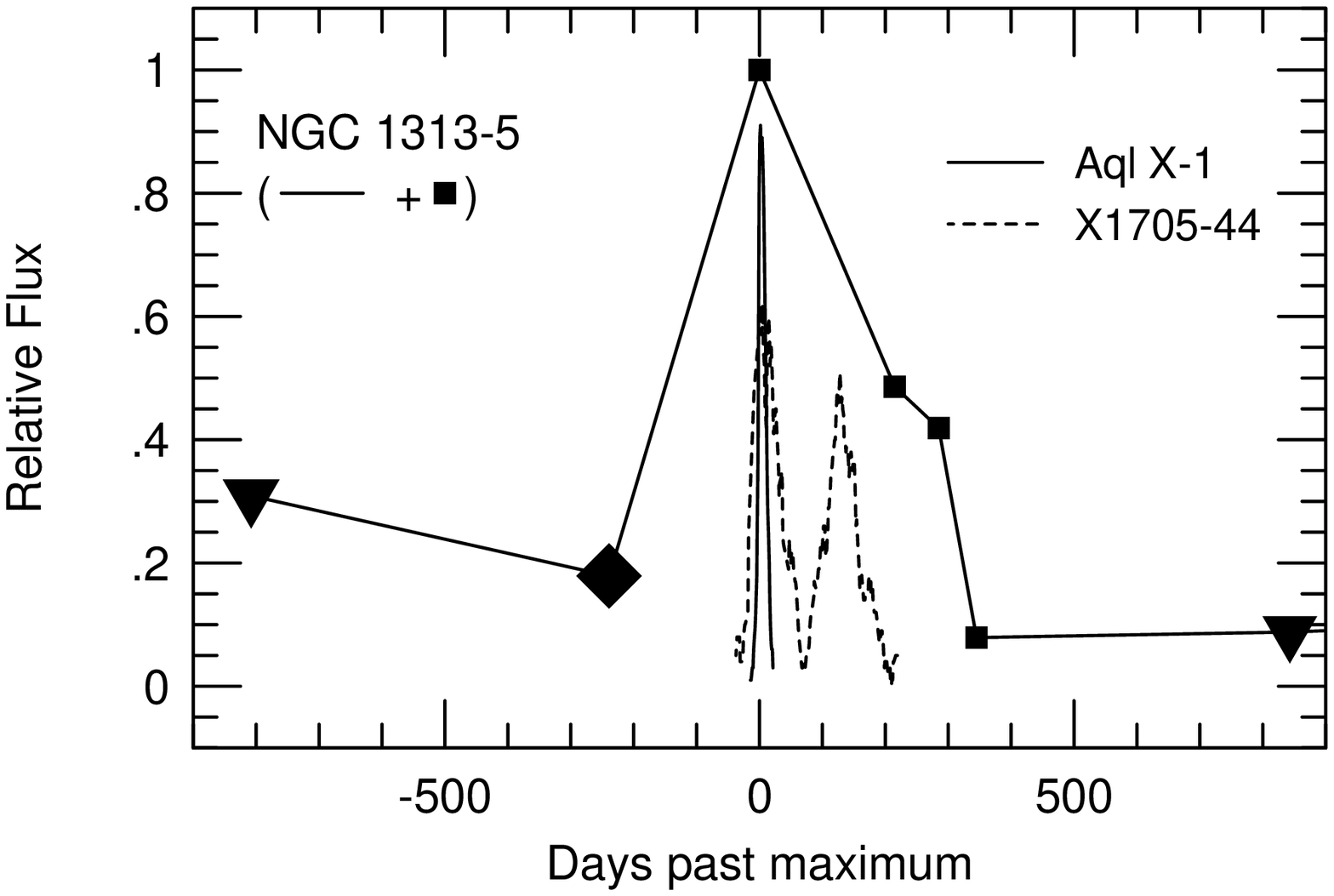}}
\end{figure*}

We arbitrarily assume the source 5 outburst occurred at or shortly
before the time of the second exposure and set that time as the time
of maximum.  We downloaded ASM data from the accumulating
database\footnote{The data are available from the {\it RXTE} WWW site
at NASA-GSFC or the {\it RXTE} ASM site at MIT.}.  We extracted times
and fluxes surrounding bursts for the sources XTEJ1550-56, GX339-4,
GRS1915+105, X0115+634, Aql X-1, and X1705-44.  We then subtracted the
minimum flux, scaled the data to the maximum, and subtracted the date
of maximum.  We smoothed the ASM data with a 5-point triangular
response to reduce the point-to-point fluctuations and aid the
visibility.  No corrections for distance or column density have been
made.  Figure~\ref{xray_nova}(a) shows the comparison between source
5's light curve and the light curves of XTEJ1550-56 and GX339-4 (top),
GRS1915+105 and X0115+634 (bottom); Figure ~\ref{xray_nova}(b) shows
the comparison to Aql X-1 and X1705-44.

The source 5 light curve clearly possesses an outburst duration of
several hundred days.  The FWHM, assuming an infinite rise at day
zero, is $\sim$200 days.  Most Galactic X-ray transients have
considerably shorter outbursts.  For example, the FWHM of the
outbursts of Aql X-1 and X0115+634 are $\sim$10 and $\sim$20 days,
respectively.  The FWZI width for Aql X-1 and X0115+634 are $\sim$30
and $\sim$50 days, respectively.  The outburst duration for source 5
is much longer, even if we missed the maximum by $\sim$100 days.

We can not eliminate the possibility that multiple outbursts occurred
in source 5 between the day 0 and day 220 HRI observations.  Another
possibility assumes the day 0 and subsequent points are the decay of a
{\it second} outburst in a multiple-outburst sequence.  The first outburst
would then fit into the unobserved $\sim$200 day gap prior to the
first detection.

If such an outburst did occur earlier, so that the second observation
fell on an exponential decay curve (as is approximately shown by the
XTEJ1550-56 curve), then source 5 must have been considerably
brighter.  By sliding the J1550 curve vertically and horizontally to a
negative ``age,'' the observed points can be made to fall at least
close to the J1550 tail of the second burst.  The FWZI is not
reproduced unless we stretch the J1550 light curve.  We have other
examples of X-ray light curves that do not require such stretching.

The light curve for GX339-4 provides a good match to the data points,
not only matching the width but also the steep decline at the end of
the burst.  If a match in light curve shape dictates physics, then by
analogy with GX339-4, source 5 is a black hole candidate.  A highly
variable source such as GRS1915+105 also can in principle provide a
match to the light curve of source 5; the data do not have the time
resolution to refute this possibility.  Interestingly, GRS1915+105 is
also a black hole candidate.

\section{Discussion}

NGC 1313 is not a particularly unusual galaxy.  Its most distinctive
attributes are a low metallicity and a high density of H II regions
along the bar \citep{Ryd+95}.  There is no evidence of nuclear
activity, nor does it show any evidence of rapid star formation, past
or present, despite having possibly interacted with its nearby
companion, NGC 1313A \citep{SB79}.

Three key issues are raised by the {\it ROSAT} data of NGC 1313:
the lack of diffuse emission, the lack of a nuclear source, and the
presence of several luminous sources.  The lack of diffuse emission
was discussed in \citet{Mi98} and will not be repeated here.
Essentially, a larger database of normal galaxies covering a broad
range of properties is needed to place the presence or absence of
diffuse emission in context.  We note, however, that low-mass galaxies
appear to contain holes in their neutral ISMs (e.g.,
\citealt{Walt99}).  The holes may provide a channel by which injected
energy escapes the ISM, rather than heating it.  While that argument
can explain the lack of a hot ISM in many dwarf galaxies, the H~I
layer in NGC 1313 does not appear to be dominated by holes (e.g.,
\citealt{Ryd+95}).

The absence of a nuclear source might indicate enhanced absorption
surrounding the nucleus.  If we assume the nuclear source is similar
to source 6, the required column density to hide its X-ray emission
and yield our observed upper limit is about 10$^{23}$ cm$^{-2}$ which
corresponds to an A$_{\rm V}$ $\sim$45, too high to be detectable in
the optical.  If enhanced absorption does explain the non-detection of
a nuclear source, an observation with {\it Chandra} or {\it XMM} may
reveal it at E $>$ 3-4 keV.

A 1 M$_{\odot}$, zero metallicity accreting source generates an
Eddington luminosity of $\sim$1.4$\times$10$^{38}$ ergs s$^{-1}$.  In
NGC 1313, excluding SN1978K, there are 5 sources above L$_{\rm Edd}$
(sources 4, 5, 6, 7, and 8).  An overestimate of the distance will
lower the luminosities; if, however, we believe the luminosities must
be sub-Eddington, the distance then must be in error by a factor of
two or more.  Given the proximity of the galaxy and the amount of
study it attracts, a distance error of that magnitude is very
unlikely.

While sources with similar luminosities to the NGC 1313 sources are
found in other nearby galaxies, for none of them does a Galactic
counterpart exist.  Sources of comparable luminosity can be found in,
for example, M51 \citep{Mars95}, M82 \citep{Coll94}, M33
\citep{Tak94}, and M101 \citep{WIP99}.  The two in M82 discussed by
Collura et al. are of particular interest, as both show strong
variability, and one has an average {\it ROSAT} band luminosity of at
least 5 $\times$10$^{39}$ ergs s$^{-1}$, making it an X-ray binary
candidate of comparable luminosity to the ones found in NGC 1313.
\citet{WIP99} discuss the number of luminous X-ray sources in several
nearby galaxies and argue that the number of sources is correlated
with the far-IR luminosity, L$_{\rm FIR}$.  L$_{\rm FIR}$ for NGC 1313
is 42.73, using the \citet{DE89} relation contained in the notes to
their Table 7 and IRAS 60 and 100 $\mu$ fluxes from the NASA
Extragalactic Database\footnote{Available from
http://nedwww.ipac.caltech.edu/}.  For NGC 1313, the L$_{\rm FIR}$
places it near the bottom of the range (slightly above M33).  The
L$_{\rm X}$ or, equivalently, the number of superluminous sources,
places NGC 1313 nearer the top of the range.  

\begin{figure*}
\caption{The HRI luminosity distribution of sources for NGC 1313
compared to the distribution for M33.  The solid and dashed lines
represent the as-observed distributions for M33 and NGC 1313,
respectively.  The dot-dashed line represents the distribution for NGC
1313 if the three most luminous sources are removed.}
\label{NL_dist}
\scalebox{0.4}{\rotatebox{-90}{\includegraphics{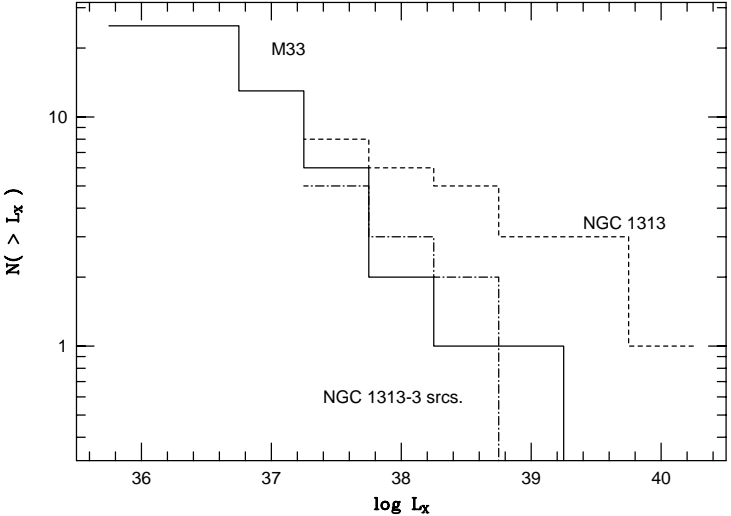}}}
\end{figure*}

If we eliminate the most luminous three objects from the luminosity
distribution, the result resembles the distribution for M33
(Figure~\ref{NL_dist}), an argument in favor of the interpretation
that sources 5, 7, 8, 9, and 10 are accreting X-ray binaries.  We can
argue that the presence of SN1978K is fortuitous; over typical
galactic times, its existence is ephemeral and happens to be visible
at present.  The other two sources, however, appear to be long-lived,
particularly the near-nuclear source.  We can ask how NGC 1313 has
produced a top-heavy luminosity distribution.  No explanation for the
``overabundance'' of very luminous X-ray sources readily presents
itself.  The existence of apparently super-Eddington non-nuclear
sources in nearby galaxies, coupled with the lack of them in our
Galaxy, is a long standing mystery in X-ray astronomy \citep{Fab98}.

\appendix

\section{Serendipitous Sources in the NGC 1313 Field}

Sources 1, 2, 3, 12, 13, 14, and 15 lie outside of the D$_{25}$ circle
centered on the galaxy.  Sources 4 and 11 = SN1978K also lie outside
that radius.  SN1978K is definitely associated with NGC 113.  The
distance to source 4 is less certain \citep{Sto94,Sto95,Ptr94a,Ptr94b}.  
We used the APM WWW site (http://www.ast.cam.ac.uk/$\sim$apmcat/) to
list potential optical counterparts.  From the APM list, we obtained a
magnitude and using the nomogram of \citet{Mac88}, estimated a log(
f$_{\rm X}$ / f$_{\rm V}$) value.  A single counterpart fell within
4$''$ of the HRI position for sources 1, 2, 12, 14, and 15.  For
sources 3 and 13, the offsets were larger (9$''$ and 7$''$,
respectively). All of the objects are fainter than mag 20.0 and all
appear to be point sources.  The nomogram indicates all are AGN or BL
Lac objects.  Table~\ref{seren_table} lists the sources, the closest
optical counterpart, its magnitude, and the nomogram.

\acknowledgements

EMS thanks Jeff McClintock for a pointer to the literature on the
light curves of X-ray binaries.  This research was supported by NASA
grant NAG5-4015 to the Smithsonian Astrophysical Observatory.

\begin{table*}
\begin{center}
\caption{{\it ROSAT} HRI Observations} \label{comp_set}
\begin{tabular}{llrrl}
         & Observation &   & Exposure & Pointing \\
Sequence &  Date  & MJD  & Time (sec) & centered on \\ \hline
400065n00\tablenotemark{a,b} & 1992 Apr 18-May 24 & 48730 &  5365 & NGC 1313 \\
600505n00\tablenotemark{a} & 1994 Jun 23-Jul 16  & 49538 & 22199 & NGC 1313 \\
500403n00 & 1995 Jan 31-Feb 10  & 49753 & 13314 & SN1978K  \\
500404n00 & 1995 Feb 2-11     & 49754 & 26898 & SN1978K  \\
600505a01 & 1995 Apr 12-20    & 49823 & 20143 & NGC 1313 \\
500403a01 & 1995 May 10-Jul 22 & 49883 & 30736 & SN1978K \\
500404a01 & 1995 May 9-Jul 21 & 49884 & 18628 & SN1978K \\
500492n00 & 1997 Sep 30-Aug 10 & 50726 & 22537 & SN1978K \\ 
500499n00 & 1998 Mar 21-Apr 20 & 50908 & 23754 & SN1978K \\ \hline
\end{tabular}
\end{center}

\tablenotetext{a}{In light curve presented in \citet{SPC96}.}

\tablenotetext{b}{Data described in \citet{Sto94} or \citet{Sto95}.}

\end{table*}

\begin{table*}
\begin{center}
\caption{Detected Sources in Summed Image}
\label{src_list}
\begin{tabular}{lcccrccl}
        & \multicolumn{2}{c}{J2000} & Total & Cts/sec & & & Old \\
 Number & RA  & Dec  & ExpT (sec) & ($\times$10$^4$) & Flux\tablenotemark{a} & L$_{\rm X}$ & Label\tablenotemark{b} \\ \hline
1 & 3:19:14.7 & -66:32:46.0 & 177241 & 12.1$\pm$1.3 & 1.3$\pm$0.2(-13) & 3.1(38) & X12\tablenotemark{c} \\
2 & 3:19:01.0 & -66:43:26.4 & 167992 & 6.9$\pm$1.0 & 7.3$\pm$0.1(-14) & 1.8(38) & new \\
3 & 3:18:34.8 & -66:44:27.1 & 169763 & 4.8$\pm$0.9 & 5.1$\pm$0.9(-14) & 1.2(38) & new \\
4 & 3:18:22.0 & -66:36:02.3 & 182585 & 149.0$\pm$3.5 & 1.6$\pm$0.04(-12) & 3.9(39) & X2 \\
5 & 3:18:21.0 & -66:30:32.3 & 182878 & 13.8$\pm$1.6 & 1.4$\pm$0.2(-13) & 3.4(38) & new \\
6 & 3:18:20.2 & -66:29:07.3 & 181387 & 515.9$\pm$5.8 & 5.4$\pm$0.06(-12) & 1.3(40) & X1 \\
7 & 3:18:18.9 & -66:32:29.9 & 183675 & 5.9$\pm$0.8 & 6.2$\pm$0.8(-14) & 1.5(38) & X8 \\
8 & 3:18:18.5 & -66:30:02.3 & 182693 & 25.5$\pm$1.7 & 2.7$\pm$0.2(-13) & 6.5(38) & X6 \\
9 & 3:18:06.0 & -66:30:12.4 & 184064 & 2.1$\pm$0.6 & 2.2$\pm$0.6(-14) & 5.3(37) & X7 \\
10 & 3:17:42.7 & -66:31:22.5 & 184757 & 2.1$\pm$0.6 & 2.2$\pm$0.6(-14) & 5.3(37) & X11 \\
11 & 3:17:38.4 & -66:33:02.5 & 182739 & 171.6$\pm$3.7 & 1.8$\pm$0.04(-12) & 4.3(39) & X3 \\
12 & 3:17:35.0 & -66:38:42.5 & 176697 & 13.7$\pm$1.2 & 1.4$\pm$0.1(-13) & 3.4(38) & X10 \\
13 & 3:17:16.4 & -66:43:47.3 & 172713 & 8.4$\pm$2.2 & 8.8$\pm$2.3(-14) & 2.1(38) & new \\
14 & 3:16:22.1 & -66:33:41.0 & 181503 & 8.1$\pm$0.9 & 8.5$\pm$0.9(-14) & 2.1(38) & X9 \\
15 & 3:16:16.8 & -66:37:40.8 & 172503 & 3.3$\pm$0.7 & 3.5$\pm$0.7(-14) & 8.4(37) & new \\ \hline
\end{tabular}

\tablenotetext{a}{Flux in the 0.1--2.4 keV band calculated assuming
unabsorbed 2 keV thermal bremsstrahlung spectrum with N$_{\rm H}$ =
6$\times$10$^{20}$ cm$^{-2}$.  This assumes all of the objects are in
NGC 1313 (D = 4.5 Mpc); for objects not in NGC 1313, this procedure
will underestimate the flux by $\sim$30\% for a source at a distance
of 1 kpc. The percent change in L$_{\rm X}$ is $\sim$-25\% for N$_{\rm
H}$ = 1.5$\times$10$^{21}$ cm$^{-2}$.}

\tablenotetext{b}{labels used in \citet{Co+95} and \citet{Mi98}.}

\tablenotetext{c}{Detected in \citet{Mi98} but not discussed; not
detected in \citet{Co+95}}
\end{center}
\end{table*}

\begin{table*}
\begin{center}
\caption{Count Rates of NGC 1313 Sources\tablenotemark{a}}
\label{src_rates}
\begin{tabular}{lcccccccc}
      & \multicolumn{8}{c}{Source Number}\\
 MJD  &      4 &      5 &      6 &      8 &      9 &     11 &     12 &
 14 \\ \hline

48370 & 18.1$\pm$2.4 & $<$1.4  & 69.8$\pm$4.0 & $<$1.4 & $<$1.4 & 13.1$\pm$2.1 & $<$1.4 & $<$1.5 \\
49538 & 9.8$\pm$ 0.9 & 4.4$\pm$0.6 & 46.5$\pm$1.6 & 1.9$\pm$0.5 & 1.5$\pm$0.6 & 15.0$\pm$1.0 & 0.8$\pm$0.6 & 0.9$\pm$0.6 \\
49753 & 22.6$\pm$1.5 & 6.7$\pm$0.9 & 64.6$\pm$2.4 & 8.5$\pm$1.1 & 4.1$\pm$1.3 & 20.2$\pm$1.5 & 2.0$\pm$1.2 & 2.6$\pm$1.3 \\
49754 & 17.2$\pm$1.0 & -- & 65.8$\pm$1.7 & -- & -- & 16.3$\pm$1.0 & -- & -- \\
49823 & 11.7$\pm$1.0 & 1.9$\pm$0.5 & 63.7$\pm$1.9 & 3.2$\pm$0.5 & $<$0.6 & 17.2$\pm$1.1 & $<$0.6 & $<$0.7 \\
49883 & 11.5$\pm$0.8 & 0.6$\pm$0.4 & 30.8$\pm$1.6 & 4.0$\pm$0.5 & 1.8$\pm$0.7 & 18.0$\pm$0.9 & 1.7$\pm$0.7 & 2.7$\pm$0.7 \\
49884 & 15.2$\pm$1.1 & -- & 36.5$\pm$1.6 & -- & -- & 18.8$\pm$1.2 & -- & -- \\
50726 & 16.5$\pm$1.0 & $<$0.4 & 71.5$\pm$1.9 & 2.8$\pm$0.6 & 1.8$\pm$0.6 & 17.3$\pm$1.1 & 2.0$\pm$0.6 & 1.6$\pm$0.6 \\
50908 & 17.3$\pm$1.1 & $<$0.4 & 33.0$\pm$1.3 & 2.3$\pm$0.5 & 1.5$\pm$0.6 & 16.8$\pm$1.1 & 0.7$\pm$0.6 & 0.8$\pm$0.6 \\

Chi-squared & 16.1 & 11.8 & 125. & 7.8 & 0.9 & 2.1 & 0.6 & 1.0 \\
Probability & $>$0.999 & $>$0.999 & $>$0.999 & $>$0.999 & 0.55 & 0.97 & 0.20 & 0.55 \\
\\ \hline
\end{tabular}
\end{center}

\tablenotetext{a}{Count rates in units of 10$^{-3}$ s$^{-1}$}
\end{table*}

\begin{table*}
\begin{center}
\caption{Description of Serendipitous Sources} \label{seren_table}
\begin{tabular}{cccccc}
         & \multicolumn{2}{c}{Closest Optical} \\
         & \multicolumn{2}{c}{Counterpart} & X-ray - Optical \\
 Source  & RA   & Dec  & Mag & offset & log(f$_{\rm X}$ / f$_{\rm V}$) \\ \hline
  1  & 03:19:16.44 & -66:32:45.9 & 0$''$.8 & 21.0 & +0.88 \\
  2  & 03:19:00.34 & -66:43:25.8 & 1$''$.9 & 20.2 & +0.30 \\
  3  & 03:18:33.55 & -66:44:27.3 & 9$''$.0 & 22.1 & +0.94 \\
 12  & 03:17:34.31 & -66:38:41.9 & 2$''$.1 & 22.2 & +1.40 \\
 13  & 03:17:16.84 & -66:43:40.9 & 7$''$.1 & 20.5 & +0.50 \\
 14  & 03:16:22.29 & -66:33:40.2 & 1$''$.5 & 20.0 & +0.30 \\
 15  & 03:16:16.66 & -66:37:37.5 & 3$''$.6 & 21.6 & +0.55 \\ \hline
\end{tabular}
\end{center}
\end{table*}


\newpage

\begin{thebibliography}

\bibitem[Bregman et al.(1993)]{Breg93} Bregman, J., Cox, C., \& Tomisaka, K. 1993, ApJ, 415, L79

\bibitem[Colbert et al.(1995)]{Co+95} Colbert, E. J., Petre, R., Schlegel, Eric M., \&
Ryder, S. 1995, ApJ, 446, 177

\bibitem[Colbert \& Mushotzky(1999)]{CM99} Colbert, E. J. \&
Mushotzky, R. M. 1999, ApJ, 519, 89

\bibitem[Collura et al.(1994)]{Coll94} Collura, A., Reale, F., Schulman, E., \& Bregman,
J. 1994, ApJ, 420, L63

\bibitem[David et al.(1997)]{Dav97} David, L., Harnden, F. R., Jr.,
Kearns, K., \& Zombeck, M. V. 1997, The {\it ROSAT} High Resolution
Imager (HRI) Calibration Report (rev. ed., Cambridge: US {\it ROSAT}
Science Data Center)

\bibitem[deVaucouleurs et al.(1991)]{deV+91} de Vaucouleurs, G., de
Vaucouleurs, A., Corwin, H. G., Buta, R., Paturael, G., \& Fouqu\'e,
P. 1991, Third Reference Catalog of Bright Galaxies (New York:
Springer-Verlag)

\bibitem[deVaucouleurs(1963)]{deV63} de Vaucouleurs, G. 1963, ApJ, 137, 720

\bibitem[Devereux \& Eales(1989)]{DE89} Devereux, N. A.. \& Eales,
S. A. 1989, ApJ, 340, 708

\bibitem[Fabbiano(1998)]{Fab98} Fabbiano, G. 1998 in The Hot Universe,
ed. K. Koyama, S. Kitamoto, \& M. Itoh (Dordrecht: Kluwer Academy
Publ), 93

\bibitem[Fabbiano \& Trinchieri(1987)]{FT87} Fabbiano, G. \&
Trinchieri, G. 1987, ApJ, 315, 46

\bibitem[Harris(1999)]{Har99} Harris, D. 1999, Bull AAS, 31:2, 719

\bibitem[Hasinger et al.(1994)]{Has94} Hasinger, G., Burg, R.,
 Giacconi, R., Hartner, G., Schmidt, M., Truemper, J., \& Zamorani,
G. 1994, A\&A, 291, 348

\bibitem[Maccacaro et al.(1988)]{Mac88} Maccacaro, T., Gioia, I.,
Wolter, A., Zamorani, G., \& Stocke, J. 1988, ApJ, 326, 680

\bibitem[Markert \& Rallis(1983)]{MR83} Markert, T. H. \& Rallis,
A. D. 1983, ApJ, 275, 571

\bibitem[Marston et al.(1995)]{Mars95} Marston, A. P., Elmegreen,
D. M., Elmegreen, B., Forman, W., \& Jones, C. 1995, ApJ, 438, 663

\bibitem[Micela et al.(1996)]{Mic96} Micela, G., Sciortino, S.,
Kashyap, V., Harnden, Jr., F. R., \& Rosner, R. 1996, ApJS, 102, 75

\bibitem[Miller et al.(1998)]{Mi98} Miller, S., Schlegel, Eric M.,
Petre, R. \& Colbert, E. J. 1998, AJ, 116, 1657

\bibitem[Peters et al.(1993)]{Pet93} Peters, W., Freeman, K.,
Forster, K., Manchester, R., \& Ables, J. 1993, MNRAS, 269, 1025

\bibitem[Petre et al.(1994a)]{Ptr94a} Petre, R., Schlegel, E., Colbert,
E., Okada, K., Makishima, K., \& Mihara, T. 1994a, in {\it New
Horizons in X-ray Astronomy}, eds. F. Makano \& T. Ohashi (Tokyo:
Universal Academy Press), 519

\bibitem[Petre et al.(1994b)]{Ptr94b} Petre, R., Okada, K., Mihara, T.,
Makishima, K., \& Colbert, E. J. 1994b, PASJ, 46, L115

\bibitem[Predehl \& Schmitt(1995)]{PS95} Predehl, P. \& Schmitt,
J. H. M. M. 1995, A\&A, 293, 889

\bibitem[Ryder et al.(1993)]{Ryd93} Ryder, S., Staveley-Smith, L.,
Dopita, M., Colbert, E. J., Petre, R., Malin, D., \& Schlegel,
E. M. 1993, ApJ, 416, 167

\bibitem[Ryder et al.(1995)]{Ryd+95} Ryder, S., Staveley-Smith, L.,
Malin, D. F., \& Walsh, W. 1995, AJ, 109, 1592

\bibitem[Sandage \& Brucato(1979)]{SB79} Sandage, A. \& Brucato,
R. 1979, AJ, 84, 472

\bibitem[Schlegel(1994)]{Sch94} Schlegel, Eric M. 1994, ApJ, 434, 523

\bibitem[Schlegel, Petre, \& Colbert(1996)]{SPC96} Schlegel, Eric M.,
Petre, R., \& Colbert, E. J. 1996, ApJ, 456, 187

\bibitem[Schlegel et al.(1999)]{Sch+99} Schlegel, Eric M. et
al. 1999, AJ, 118, 2689

\bibitem[Schlegel, Finkbeiner, \& Davis(1998)]{SFD98} Schlegel, D. J.,
Finkbeiner, D., \& Davis, M. 1998, ApJ, 500, 525

\bibitem[Snowden(1998)]{Sno98} Snowden, S. L. 1998, ApJS, 117, 233

\bibitem[Stocke et al.(1995)]{Sto95} Stocke, J., Wang, Q. D.,
Perlman, E., Donahue, M., \& Schachter, J. 1995, AJ, 109, 1199

\bibitem[Stocke et al.(1994)]{Sto94} Stocke, J., Wang, Q., Perlman,
E., Donahue, M., \& Schachter, J. 1994, in The Soft X-ray Cosmos,
ed. E. Schlegel \& R. Petre (New York: AIP), 75

\bibitem[Takano et al.(1993)]{Tak94} Takano, M., Mitsuda, K.,
Fukazawa, Y., \& Nagase, F. 1993, ApJ, 436, L47

\bibitem[Tully(1988)]{Tul88} Tully, R. 1988, Nearby Galaxies Catalog,
(Cambridge: Cambridge University Press)

\bibitem[Walter(1999)]{Walt99} Walter, F. 1999, PASA, 16, 106

\bibitem[Wang, Immler, \& Pietsch(1999)]{WIP99} Wang, Q. D., Immler,
S., \& Pietsch, W. 1999, ApJ, 523, 121

\bibitem[White, Nagase, \& Parmar(1995)]{WNP95} White, N., Nagase,
F., \& Parmar, A. 1995 in X-ray Binaries, ed. W. Lewin, (Cambridge:
Cambridge University Press), 1

\end{thebibliography}
\end{document}